\documentclass[a4paper,11pt]{article}

\usepackage{cite}\usepackage{amsmath}
\usepackage{amssymb}
\usepackage{graphicx, color}
\usepackage{palatino}
\newcommand{\sups}[1]{$^{#1}$}

\makeatletter

\def\@biblabel#1{#1.}

\makeatother

\begin{document}

\begin{titlepage}

\vspace{20 pt}

\noindent\textsf{\textbf{\Large A condensation-ordering mechanism in\\ nanoparticle-catalyzed peptide aggregation}}

\vspace{20 pt}

\noindent\textsf{\textbf{Stefan Auer\sups{1,\ast}, Antonio Trovato\sups{2}, and Michele Vendruscolo\sups{3}}}

\date{\today}

\pagenumbering{arabic}

\vspace{80 pt}

\noindent\sups{1}\textsl{Centre for Self Organising Molecular Systems, 
University of Leeds, Leeds LS2 9JT, UK}

\vspace{10 pt}

\noindent\sups{2}\textsl{Dipartimento di Fisica, Universit\'a di Padova, CNISM and INFN, Via Marzolo 8, 35131 Padova, Italy}

\vspace{10 pt}

\noindent\sups{3}\textsl{Department of Chemistry, Lensfield Road, Cambridge,
CB2 1EW, UK} \\

\vspace{10 pt}
\noindent\sups{\ast} \textsl{Corresponding Author
(\texttt{s.auer@leeds.ac.uk}).}

\vspace{60 pt}

\end{titlepage}

\section*{Abstract}
Nanoparticles introduced in living cells are capable of strongly
promoting the aggregation of peptides and proteins.  We use here
molecular dynamics simulations to characterise in detail the process
by which nanoparticle surfaces catalyse the self-assembly of peptides
into fibrillar structures.  The simulation of a system of hundreds of
peptides over the millisecond timescale enables us to show that the
mechanism of aggregation involves a first phase in which small
structurally disordered oligomers assemble onto the nanoparticle and a
second phase in which they evolve into highly ordered $\beta$-sheets
as their size increases.
\vspace{10truemm}

\section*{Author Summary}
Protein misfolding and aggregation are associated with a wide variety
of human disorders, which include Alzheimer's and Parkinson's diseases
and late onset diabetes. It has been recently realised that the
process of aggregation may be triggered by the presence of
nanoparticles.  We use here molecular dynamics simulations to
characterise the molecular mechanism by which such nanoparticles are
capable of enhancing the rate of formation of peptide aggregates.  
Our findings indicate that nanoparticle surfaces act as a catalyst
that increases the local concentration of peptides, thus facilitating
their subsequent assembly into stable fibrillar structures.
The approach that we present, in addition to providing a description
of the process of aggregation of peptides in the presence of nanoparticles,
will enable the study of the mechanism of action of a variety of other potential
aggregation-promoting agents present in living organisms, 
including lipid membranes and other cellular components.

\clearpage

\section*{Introduction}

With the advent of nanoscience much interest has arisen about the ways
in which nanoparticles interact with biological systems, because of
their potential applications in nanotechnology and effects on human
health \cite{klein07,colvin07,fischer07,lynch08,schulze08,lewinski08,sanvicens08}.
When nanoparticles are introduced in a living organism they may
interact with a variety of different cellular components with yet
largely unknown pathological consequences.  These concerns have been
articulated particularly in the case of misfolding disorders with
increasing evidence, for example, about an association between
exposure to heavy metals and an enhanced risk of developing
Parkinson's disease \cite{uversky01a}.  Such misfolding diseases are
caused by the aberrant association of peptides and
proteins \cite{chiti06}, which result in fibrillar aggregates that
share a common cross-$\beta$ structure of intertwined layers of
$\beta$-sheets \cite{chiti06}.  Although is well known that such
aggregates are formed in a nucleation-dependent
manner \cite{jarrett93,chiti06} and that very often nucleation
phenomena are known to be triggered by external factors \cite{sear07},
experimental reports on protein aggregation in heterogeneous systems
have only begun to
emerge \cite{linse07,cedervall07,cabaleiro-lago08,lundqvist08}.  These
studies are important, since peptides and proteins {\em in vivo} often
interact with a variety of potential seeding agents such as
macromolecular complexes or membranes, which may strongly influence
their aggregation behaviour.  Indeed, it is well known that
colloids \cite{linse07,lundqvist08,cabaleiro-lago08}, lipid
bilayers \cite{knight04}, and liquid-air, liquid-solid or liquid-liquid
interfaces \cite{powers01,lu03} can have significant effects in
promoting amyloid formation.  It has also been recently shown that,
{\em in vivo}, nanoparticles are often covered by peptides and
proteins that determine their behaviour in the
cell \cite{cedervall07,lundqvist08}.  Despite these observations, the
detailed processes underlying the association of proteins on surfaces
or nanoparticles have so far remained elusive.

In this work we use molecular dynamics simulations to investigate the
molecular mechanism of peptide self-assembly in the presence of
spherical nanoparticles. Although computational studies using full
atomistic models have provided considerable insight into the role of
fundamental forces in promoting the self-assembly of polypeptide
chains, they are restricted to relatively small systems of peptides and short
timescales \cite{ma02,hwang04a,buchete05,hills07,nguyen07,cheon07,cheon08,baumketner07,baumketner08,tarus08,vitalis08}.
Coarse-grained models have proven capable of following the evolution
of systems composed of larger numbers of peptides over
longer timescales. The most tractable models are confined to a
lattice \cite{dima02,li08,zhang09}, although in these cases
the structural details used to represent polypeptide chain conformations
are necessarily limited. Off-lattice protein
models used to simulate protein aggregation include two-state
models in which the protein
can adopt, in addition to a native state, a state that is prone to
$\beta$-sheet formation \cite{pellarin06,pellarin07}, 
two-bead models, in which each amino acid is
represented by two spheres with a knowledge-based
potential \cite{peng04}, and fine-grained models with explicit
representation of the side chains in combination with a
phenomenological force
field \cite{nguyen04,urbanc04,bellesia07,derreumaux07}. The more
detailed is the protein model, the higher is the computational cost and the
larger is the number of parameters required to specify the force
field \cite{tozzini05}. The studies mentioned above
have investigated the process of protein self-assembly in homogeneous systems in which
external factors such as nanoparticles or other molecules are
absent. Only very recently, Friedman {\em et al.} 
investigated the process of assembly of amphiphatic peptides 
in the presence of lipid vescicles \cite{friedman09}.

In the present work, we adopted an off-lattice protein
model \cite{hoang04,banavar04,hoang06,banavar07}, in which the protein
backbone is represented as tube that embeds a chain of C$_{\alpha}$ atoms subject to
interactions that are common to all polypeptide chains, including
excluded volume constraints, hydrophobic attractions, bending
rigidity, and cooperative hydrogen bonds (see Materials and
Methods). The major strength of the model is its ability to reproduce
rather accurately secondary structure elements through the excluded-volume
effects due to the tube geometry \cite{hoang04,banavar04,hoang06,banavar07}, 
which enables the use of a relatively
simple force field and thus is very efficient computationally. 
By using this type of model, we already provided
insight into the early stages of the aggregation process, to establish
the existence of a general condensation-ordering transition for
protein aggregation \cite{auer07,auer08}, and to reveal a 
self-templated nucleation mechanism \cite{auer08a} that is able to explain
a key feature observed in protein aggregation - the coupling between
the initial formation of oligomeric assemblies and their subsequent
rearrangement into a highly ordered cross-$\beta$ structures. In this
work, we show the feasibility of simulating hundreds of peptides over
several milliseconds, and we characterise in detail the molecular
mechanism of self-assembly of the peptides at the surface of
nanoparticles. This process takes place in two steps - at first the
peptides associate on the surface thus increasing their local
concentration and subsequently they undergo a process of reordering
into $\beta$ sheet structures, which is driven by the tendency to form
hydrogen bonds.

\section*{Results and Discussion}

\begin{figure}
\begin{center}
\includegraphics[width=16cm]{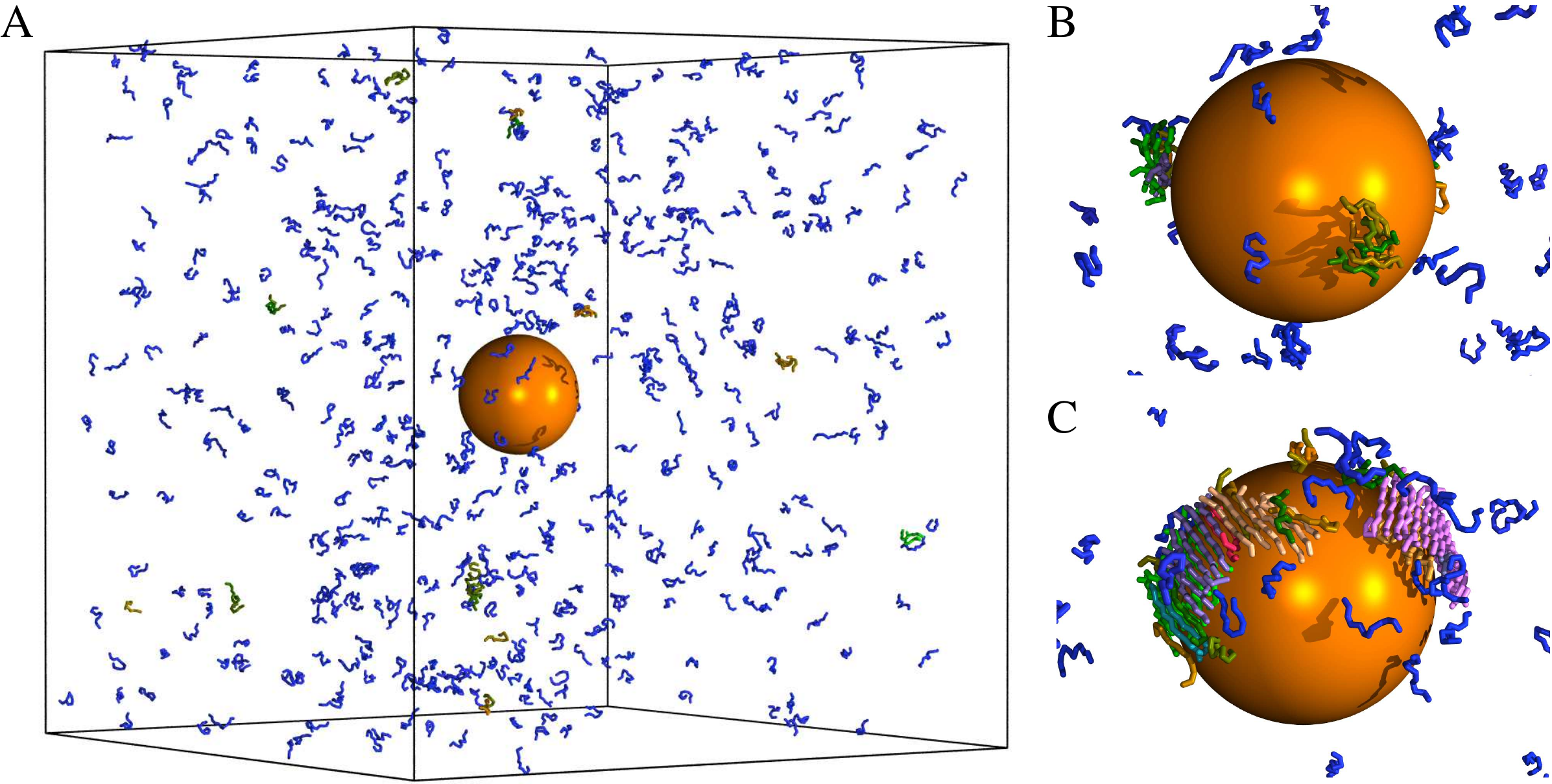}
\caption{ {\bf Illustration of the ``condensation-ordering'' mechanism
of peptide self-assembly in the presence of a hydrophobic
nanoparticle}. (A) Initially, at $t=3.9$ microseconds, the
peptides are in their monomeric state. (B) At intermediate
times, $t=0.195$ milliseconds, small oligomeric assemblies form on the
nanoparticle surface. (C) At later times, $t=0.78$ milliseconds,
these oligomers re-order into fibrillar structures as their size
increases. Peptides that do not form intermolecular hydrogen bonds are
shown in blue, while peptides that form intermolecular hydrogen bonds
are assigned a random colour, which is the same for peptides that
belong to the same $\beta$-sheet. Two peptides are defined as
belonging to the same cluster if their centres of mass distance is
less than $5$\AA. Two peptides are taken to participate within a
$\beta$-sheet if they form more than four inter-chain hydrogen bonds
with each other.  The spherical nanoparticle is displayed in orange in
the centre of the simulation box; the diameter of the peptides is
slightly reduced for illustration purposes.  Panels (B) and (C) show
enlarged views of the nanoparticle-peptide system.  The simulation was
performed at $c=3.4mM$, $T^*/T_f^*=1.1$, $\sigma=110\AA$, and
$e_{HP,S}/e_{HP}=2$.
\label{fig1}}
\end{center}
\end{figure}

Starting from the experimental observation that amyloid formation is a
phenomenon common to most polypeptide chains \cite{chiti06}, and that
systems of polyamino acids have been shown to form amyloid
assemblies \cite{aggeli01,fandrich02}, we investigate the aggregation
behaviour of $512$ 12-residue polyamino acids in the presence of
spherical nanoparticles (Fig.~\ref{fig1}A) as a model system to reveal
the general properties of this phenomenon.  The peptides that we
considered exhibit an $\alpha$-helical native structure below their
folding temperature $T_f^*=0.63$ (expressed here in reduced units, see
Materials and Methods) and an extended random coil structure above it.
The aggregation behaviour that we observe depends on the diameter
$\sigma$ of the spherical nanoparticle and on the strength of the
peptide-nanoparticle interaction, $e_{HP,S}$, which is the energy
gained when the distance between a C$_{\alpha}$ atom representing the
peptide molecules and the nanoparticle surface is smaller than
$10\AA$. This range was chosen to be similar to that of the pairwise
hydrophobic inter-residue interaction, since both interactions are
effectively due to hydrophobic solvation effects.

We first performed molecular dynamics simulations at a peptide
concentration, $c=3.4mM$, and reduced temperature, $T^*=0.69$, at
which in absence of the nanoparticle aggregation does not occur on the
timescale accessible to the simulation.  At this temperature most
peptides are unfolded since $T^*/T_f^*=1.1$.  The nanoparticle
diameter is set to $\sigma=110\AA$ and the interaction strength
between the nanoparticle and the peptides, $e_{HP,S}/e_{HP}=2$, is set
to twice the value of the hydrophobic attraction, $e_{HP}$, between
different C$_{\alpha}$ atoms.  The results of a representative
molecular dynamics trajectory in presence of this weakly hydrophobic
nanoparticle are shown in Fig.~\ref{fig1}.

\begin{figure}
\begin{center}
\includegraphics[width=16cm]{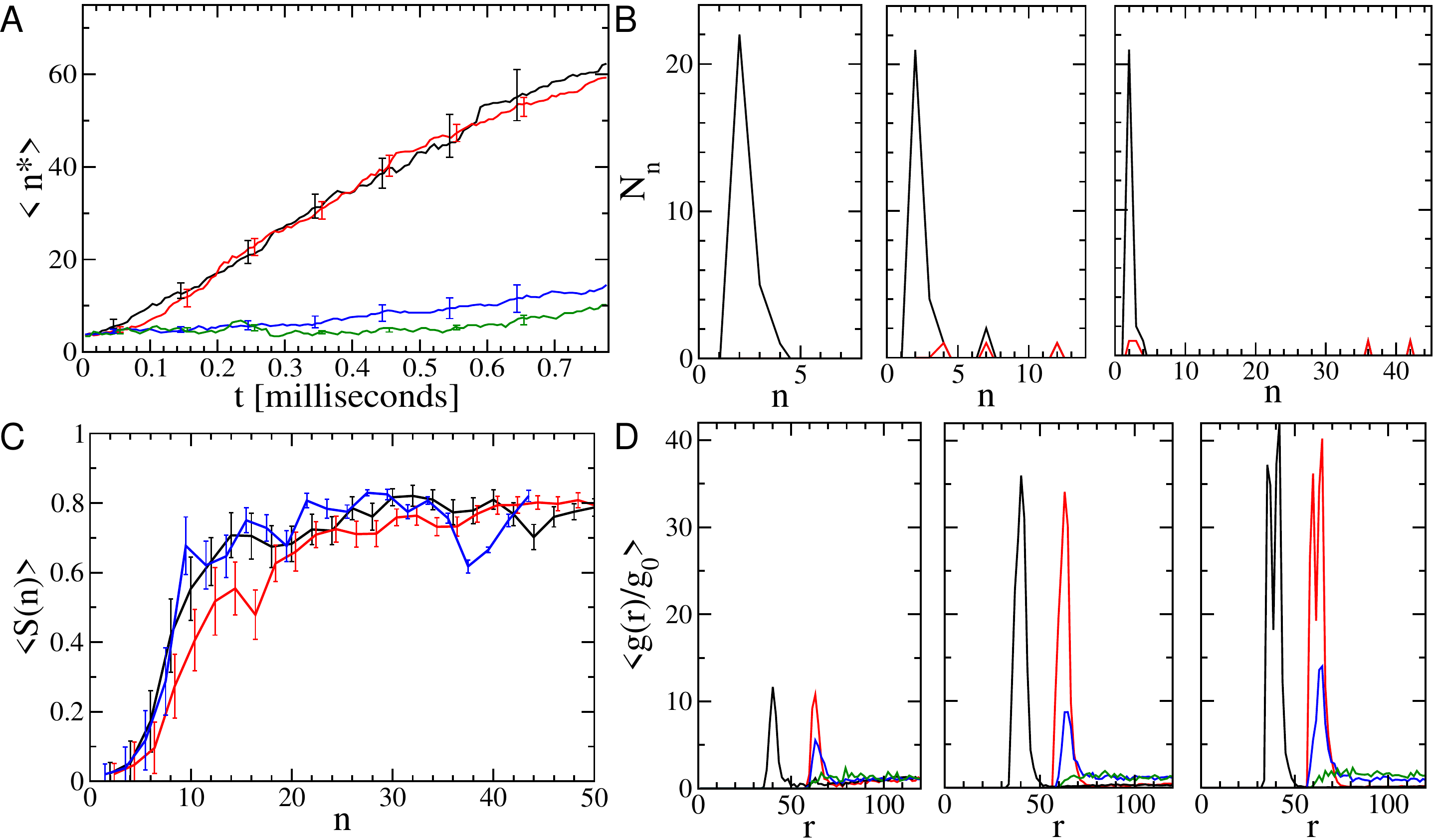}
\caption{{\bf Structural analysis of the nanoparticle-induced
  self-assembly mechanism}. (A) Average size of the largest cluster
  $n^*$ observed during a simulation in presence of a hard sphere
  nanoparticle: $\sigma = 110\AA$, $e_{HP,S}=0$ (green line), and
  several hydrophobic nanoparticles that differ in diameter and
  hydrophobicity: $\sigma = 110\AA$, $e_{HP,S}/e_{HP}=1$ (blue line),
  $\sigma = 110\AA$, $e_{HP,S}/e_{HP}=2$ (red line), and $\sigma =
  65\AA$, $e_{HP,S}/e_{HP}=2$ (black line). The results are averaged
  over ten independent simulation runs and the error bars correspond
  to the standard deviation of the mean. (B) Number of clusters $N_n$
  of size $n$ as a function of time: $t=3.9$ microseconds (left),
  $t=0.195$ milliseconds (middle), $t=0.78$ milliseconds (right), for
  the MD trajectory and parameters described in Fig.~\ref{fig1}. Black
  lines correspond to all clusters formed in the system; red lines
  correspond to the number of clusters formed on the nanoparticle
  surface. (C) Structural order parameter $S$ as a function of the
  cluster size $n$ averaged over ten independent simulations. The line
  colours are as described in (A). (D) Normalized density profile
  $g(r)/g_0$, where $g_0$ is the bulk density of the system, as a
  function of the distance from the centre of mass of the nanoparticle
  at the beginning of the simulation, $t=3.9$ microseconds (left
  panel), intermediate times, $t=0.195$ milliseconds (middle panel),
  and at the end $t=0.78$ milliseconds (right panel). The different
  line colours are as described in (A)~\label{fig2} and correspond to
  the different seed sizes and peptide seed interaction energies. The
  results are averaged over ten independent simulations and the error
  bars correspond to the standard deviation of the mean.}
\end{center}
\end{figure}

We consistently observe that the presence of this hydrophobic
nanoparticle effectively removes the lag-time prior to aggregation
(Fig.~\ref{fig2}A, red line) by triggering the condensation of
peptides on the nanoparticle surface to initially form small
disordered oligomers (Fig.~\ref{fig1}B), which re-order into
$\beta$-sheets as their size increases (Fig.~\ref{fig1}C). Although
dimers and trimers constantly form and dissolve throughout the
simulation (Fig.~\ref{fig2}B), larger oligomers appear only on the
nanoparticle surface and at later times. For example, at $t=0.195$
milliseconds, we observed one cluster of size $n=7$ in solution, and
two clusters of sizes $n=7$ and $n=12$ on the nanoparticle surface
(see Figs.~\ref{fig1}B,~\ref{fig2}B, middle panel). At the end of the
simulation, at $t=0.78$ milliseconds, the two oligomers on the
nanoparticle surface had grown to sizes $n=36$ and $n=42$, whereas the
oligomers in solution dissolved (see Fig.~\ref{fig2}B, right panel).
Animations representing the molecular dynamics trajectory
corresponding to Fig.~\ref{fig1} (Videos S1 and S2, Supplementary Material) also
illustrate that the small oligomers on the seed surface can diffuse
rather freely, and that two of them collide and merge into a larger
one.

In order to provide a detailed analysis of the structure of the
clusters that form on the nanoparticle surface we calculated the
liquid crystalline order parameter $S= \langle
3\cos^2(\theta)-1/2\rangle$, a measure for the alignment between
different strands in a single $\beta$-sheet and between different
$\beta$-sheets, where $\theta$ is the angle between neighbouring
peptides within an aggregating cluster.  Our calculations confirm that
small clusters are disordered and only larger ones reach a value
$S\simeq 0.8~$ characteristic for liquid crystalline ordering
(Fig.~\ref{fig2}C).  To investigate the effect of the nanoparticle on
the local structure around its surface we calculated the density
profile of peptides as a function of their distance from the centre of
mass of the nanoparticle. Our results illustrate that the presence of
a hydrophobic nanoparticle leads to the formation of a high density
shell at the nanoparticle surface, which becomes more pronounced as
the simulation progresses (Fig.~\ref{fig2}D). The enhanced density of
peptides at the nanoparticle surface increases the probability to form
clusters, which will eventually trigger the formation of small clusters
on the seed surface. The appearance of a double layer structure seems
to be dependent on the reaching of a local density threshold (compare
different curves in ~\ref{fig2}D, rightmost panel). The facilitation
of the nucleation step within an intermediate dense assembly is well
known in crystallisation \cite{tenwolde97} and was also observed in the
assembly of peptides into cross-$\beta$ structures \cite{auer08}. These
results also provide a molecular illustration of the dynamics of
polypeptide chains associated with the ``corona'' effect observed in
recent experiments \cite{cedervall07,lundqvist08}, which has revealed
that {\em in vivo} nanoparticles are always covered by biological
molecules.  

In our simulations the lag time for nanoparticle induced peptide
aggregation is about a microsecond which is quite short compared to
the lag times typically observed in experiments. The latter range from
some hours to several days, but it should be noted that both peptide
concentration ($3.4 mM$) and, especially, nanoparticle concentration
($6.5\mu M$) are much higher than in experiments ($40-80 \mu M$ and
$4-90 pM$, respectively) \cite{linse07}. This implies that in our
simulations nucleation barriers are essentially removed by the
nanoparticle whereas in experiments they are still high.

\begin{figure}
\begin{center}
\includegraphics[width=16cm]{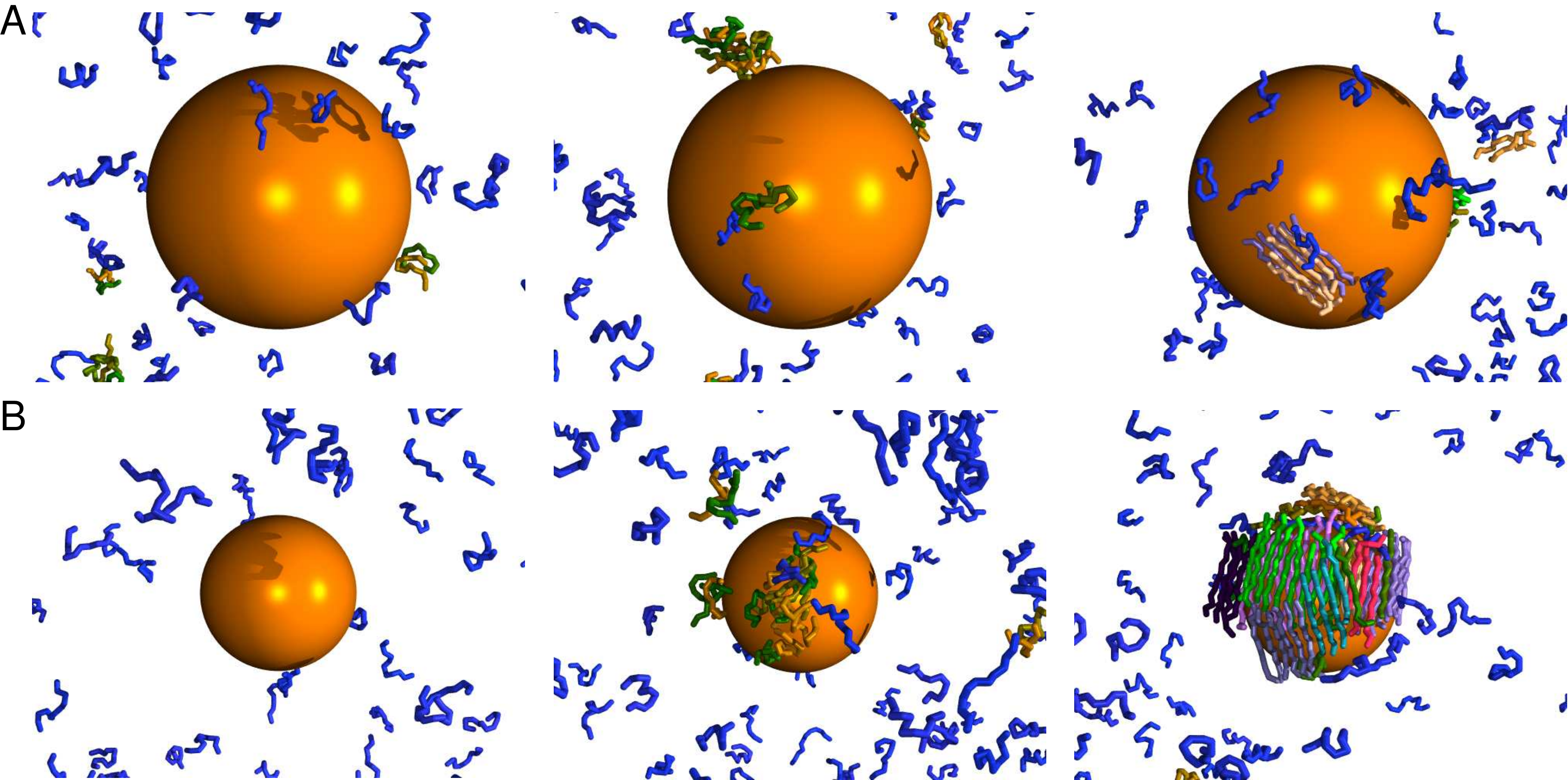}
\caption{{\bf Illustration of the condensation-ordering mechanism for
    different hydrophobicity of the nanoparticle, nanoparticle
    diameter.} (A) $\sigma=110\AA$, $e_{HP,S}/e_{HP}=1$ at
    $t=3.9$ microseconds(left), $t=0.165$ milliseconds (middle),
    $t=0.78$ milliseconds (right). (B) $\sigma=65\AA$,
    $e_{HP,S}/e_{HP}=2$ at $t=3.9$ microseconds (left), $t=0.195$
    milliseconds (middle), $t=0.78$ milliseconds (right). The
    concentration and temperature are $c=3.4$mM, $T^*/T_f^*=1.1$
    respectively, and the colour code is as described in
    Fig~\ref{fig1}.
\label{fig3}}
\end{center}
\end{figure}

In our model system the binding of peptides to the nanoparticle is
stronger for the more hydrophobic surface (Fig. ~\ref{fig2}D). As a
result, increasing the hydrophobicity of the nanoparticle reduces the
lag time prior to aggregation (Figs. ~\ref{fig2}A and
~\ref{fig3}). The same correlation between a stronger
nanoparticle-protein binding and a more enhanced reduction of the lag
time prior to aggregation is found in experiments on
$\beta2$-microglobulin fibrillation in the presence of hydrophobic
copolymer nanoparticles \cite{linse07}. This supports the already
suggested notion that a hydrophobic nanoparticle favors aggregation by
leading to a local increase in peptide concentration around its
surface. Note that, as $\beta2$-microglobulin is found to bind more
weakly to the more hydrophobic nanoparticles, the latter are found to
be {\em less} effective as well in reducing aggregation lag
times \cite{linse07}. Both these facts are not reproduced in our
simulations, reflecting our most simple modeling of the hydrophobic
effect and of the internal structure of both the nanoparticle and the
protein.

We did not observe an increase of the lag time prior to aggregation by
using a smaller nanoparticle diameter, $\sigma=60\AA$. The
fluctuations in the size of the largest cluster are nevertheless
larger, indicating that the bigger nanoparticle is a slightly more
efficient seed. Experimentally, it was shown that curvature effects can
strongly affect the fibrillation kinetics in a way which depends on
solution conditions \cite{linse07}. In our simulations we simply
cannot observe this effect because the nucleation barriers are
effectively removed by the nanoparticle.

As a final remark, we observe that the molecular mechanism associated
with the condensation ordering transition for peptide nanoparticle
association described here is independent of particle size and
hydrophobicity. The structural reorganization of protein chains in the
early disordered oligomeric assemblies from their native or
unstructured conformation to the cross-$\beta$ state may be more easily
observed by experiments using a nanoparticle as it localizes the
nucleation event, which may enable to monitor the reorganization process 
by fluorescence methods.

\section*{Conclusions}

We have characterised the process of nanoparticle-catalysed peptide
aggregation in terms of a condensation-ordering mechanism and
investigated its dependence on the nanoparticle diameter and the
strength of the nanoparticle-peptide interactions. A similar
mechanism of aggregation has already been observed in the absence of
catalysing factors \cite{auer07,auer08,auer08a,serio00}, suggesting
that the process of aggregation is driven in both cases by the
intrinsic tendency of polypeptide chains to associate by forming ordered
networks of hydrogen bonds \cite{dobson03a,knowles07}. In the case
that we have studied here, the initial condensation of peptides is
initiated by nanoparticle surfaces to form small disordered
oligomeric structures that subsequently re-order into $\beta$-sheets
as their size increases.  Although this mechanism will be modulated by
specific sequence-dependent interactions for more complex amino acid
sequences, our findings are consistent with recent experiments on
seeded fibrillation \cite{linse07}. These results therefore suggest
that the process of protein aggregation can be speeded up by the
presence of factors capable of increasing the local concentration of
proteins and thus promoting the formation of disorder oligomeric
assemblies whose presence in turn facilitates the conversion of
soluble proteins into highly ordered fibrillar structures.

\section*{Materials and Methods}

\subsection*{Description of the model}

We used a modified version of the tube model \cite{hoang04}.  In this
model, residues are represented by their $C_\alpha$ atoms, which are
connected into a chain with a distance of $3.8\pm 0.2$\AA\ between
neighbouring atoms. The tube geometry is approximated by assigning a
diameter of $3.8\AA$ to the $C_\alpha$ atoms. Neighbouring
$C_{\alpha}$ atoms are not allowed to interpenetrate. Bond angles are
restricted between $82^\circ$ to $148^\circ$, and an analogue of
bending rigidity is introduced by means of an energetic penalty,
$e_{\rm S} >0$, for values of bond angles lower than $107.15^\circ$;
these are the same criteria used previously \cite{hoang04}. The
introduction of $e_{\rm S}$ is useful to mimic the constraints placed
on local conformations by the presence of side chains, as usually
visualised by Ramachandran plot. Hydrophobicity enters through a
pairwise-additive interaction energy of $e_{\rm HP}$ (positive or
negative) between any pair of residues $i$ and $j>i+2$ that approach
closer than $7.5$\AA.

\begin{figure}
\begin{center}
\includegraphics[width=16cm]{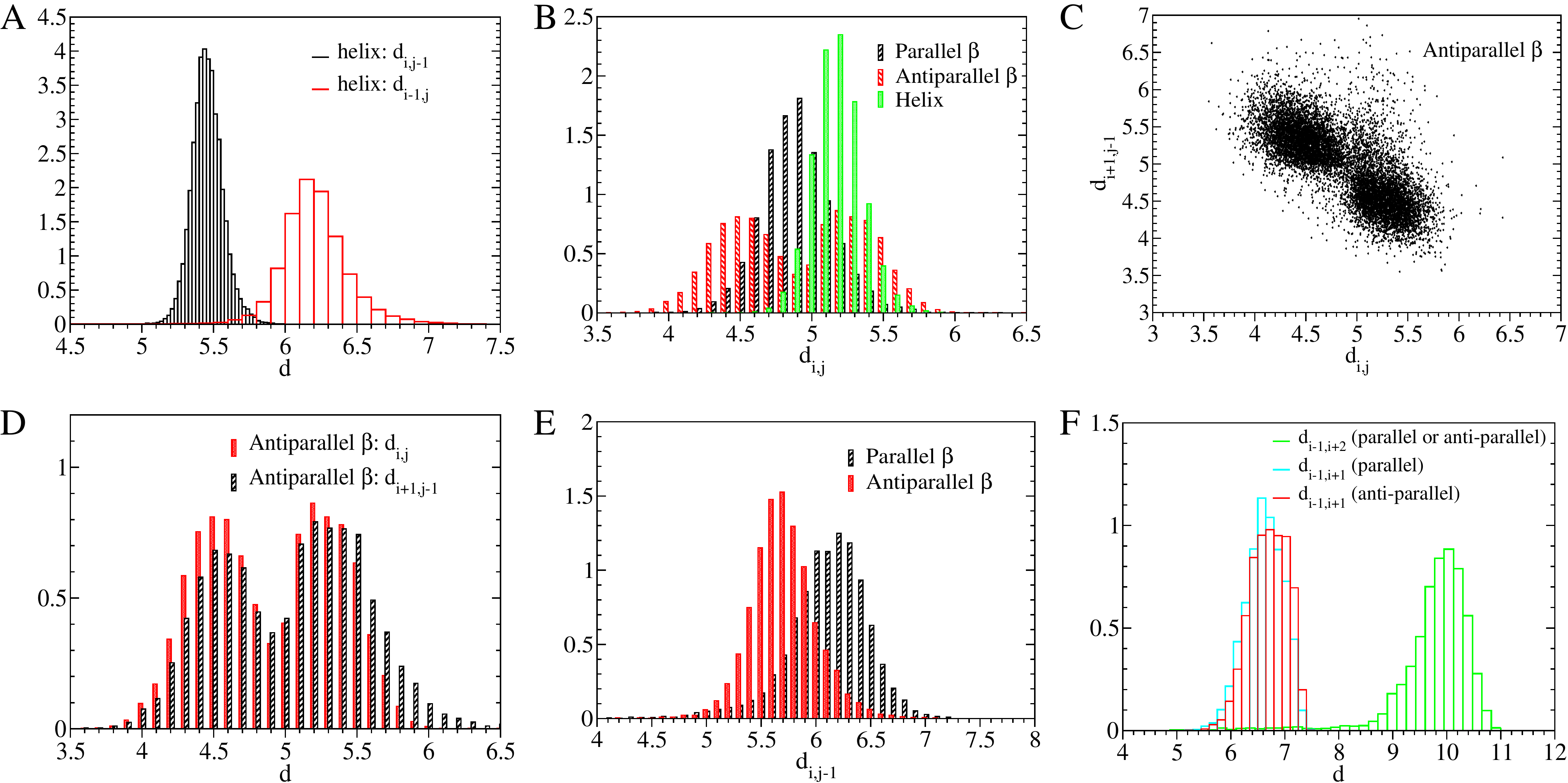}
\caption{{\bf Normalised histograms of distances between C$_{\alpha}$ atoms
  involved in backbone-backbone hydrogen bonds.} The analysis is based on
  $500$ high resolution PDB structures \cite{lovell03} and
  used to define hydrogen bonds in the protein model employed in our
  simulation. (A) Histogram of distances $d_{i,j-1}$ and
  $d_{i-1,j}$ between $C_{\alpha}$ atoms $(i,j-1)$ and $(i-1,j)$ used
  to define a $\alpha$ helical hydrogen bond assigned to atoms
  $(i,j)$ with $j=i+3$. (B) Histogram of distances $d_{i,j}$ between
  $C_{\alpha}$ atoms $(i,j)$ that form a parallel, anti-parallel, or
  helical hydrogen bond. For the $\alpha$ helical hydrogen bond
  $j=i+3$. (C) Illustration of the alternation of distances for
  consecutive $C_{\alpha}$ atoms that form anti-parallel hydrogen
  bonds. (D) as in (C). (E) The distance $d_{i,j-1}$ is
  used to define hydrogen bonds between atoms $(i,j)$ in parallel and
  anti parallel $\beta$-sheets. (F) Distances used to define
  cooperative hydrogen bonds between two consecutive atoms $(i,j)$ and
  $(i+1,j+1)$ that form parallel $\beta$-sheets or $(i,j)$ and
  $(i+1,j-1)$ that form anti parallel $\beta$-sheets.\label{Fig4}}
\end{center}
\end{figure}

The quasi-cylindrical symmetry of the tube is broken by the geometric
requirements of hydrogen bonds. These geometrical requirements were
deduced from an analysis of $500$ high resolution PDB native
structures \cite{lovell03}, from which we computed the normalised
histograms of distances between C$_{\alpha}$ atoms involved in
backbone-backbone hydrogen bonds which are shown in Fig.~\ref{Fig4}.
The distances we used to define the hydrogen bonds at the $C_{\alpha}$
atom level are summarised in Table~\ref{Tab1}. Our definitions
distinguish between hydrogen bonds that belong to a $\alpha$ helix,
parallel or anti-parallel $\beta$ sheets. We emphasise the fact that
there is not a full correspondence with the real hydrogen bonds formed
between amide and carboxyl backbone groups. For instance, there are
two different kinds of residue pairs facing each other in nearby
anti-parallel $\beta$-strands. In the first kind, the two hydrogen
bonds are formed between the two residues, whereas in the second kind,
no hydrogen bond is formed between them. The two kinds alternate along
the pair of nearby strands. In our definition of hydrogen bonds based
on $C_{\alpha}$ atoms, we will say that for both kind of pairs one
hydrogen bond is formed between the two $C_{\alpha}$'s. Yet, we keep
track of the peculiar geometry of hydrogen bonds within anti-parallel
$\beta$-sheets by using two different sets of distances, which we call
anti-parallel $1$ and anti-parallel $2$, as the distances between
consecutive $C_{\alpha}$ pairs facing each other on nearby
$\beta$-strands do indeed alternate.  Furthermore, we request that one
residue cannot form more than two hydrogen bonds, and that the first
and last $C_{\alpha}$ atoms of a peptide do not at all.  Hydrogen
bonds may form cooperatively between residues $(i,j)$ and $(i+1,j+1)$
[or $(i,j)$ and $(i+1,j-1)$ for anti-parallel hydrogen bonds], thereby
gaining an additional energy of $0.3e_{\rm HB}$. The distance criteria
for cooperative hydrogen bonds within $\beta$-sheets are obtained from
Fig.~\ref{Fig4}F and summarised in Table~\ref{Tab1}.

The energy of hydrogen bonds was set to $e_{\rm HB}=-3kT_o$, where
$kT_o$ is the thermal energy at room temperature and $k$ is the
Boltzmann constant.  This energy correspond to the experimental one
($1.5$ kCal/mol at room temperature \cite{fersht85}). Values of the
hydrophobicity and stiffness parameters, $e_{\rm HP}$ and $e_{\rm S}$,
are given in units of $kT_o$ and the reduced temperature is
$T^*=T/T_o$.  In all our simulations we set $e_S=0.9kT_o$ and
$e_{HP}=-0.15kT_o$. The ratio of a hydrogen bonding energy to
hydrophobic energy is $e_{\rm HB}/e_{\rm HP}=20$.  As the number of
hydrophobic contacts within an oligomer is usually about one order of
magnitude larger than the number of hydrogen bonds, our choice ensures
that these interactions provide similar contributions to the potential
energy of the oligomer \cite{auer08}. For this set of model parameter
the peptide folds into a native $\alpha$-helical state below the
folding temperature $T_f^*=0.63$.  $e_{HP,S}$ is the parameter which
determines the strength of the interaction energy between $C_{\alpha}$
atoms representing the peptide molecules and the seed particle. The
range of the peptide seed interaction is set to $10\AA$ from the
nanoparticle surface.

\subsection*{Simulation techniques}

We performed discontinuous molecular dynamics (DMD)
simulations \cite{alder59}, which is a fast alternative to standard
molecular dynamics simulations.  The main difference is that in DMD
simulations the system evolves on a collision by collision basis, and
requires the calculation of the collision dynamics and the search for
the next collision. In the simulations we used a cubic box, of side
$633$\AA, and applied periodic boundary conditions. The implementation
of our definition for the hydrogen bonding requires some additional
consideration. In order to prevent that one residue forms three
hydrogen bonds we treat the associated collision as fully elastic. In
order to implement and consider cooperative hydrogen bonding we keep
and update a list of all hydrogen bonds formed in the system at all
times. Note that a recalculation of the hydrogen bonds formed in the
system without considering this list can lead to a different
result. Independent starting configurations were generated at
$T^*=0.75$ and rapidly cooled down to $T^*=0.69$ at the beginning of
each simulation run. We performed all our simulation in the NVT
ensemble using an Anderson thermostat.

In order to associate the number of collision steps performed in our
simulation to a real time we measured the long time self-diffusion
coefficient of our model peptide, $D_{\rm pep}=0.0085\AA^2/$(collision
step), and matched it to experimental data.  We took from the
literature the value for the self-diffusion coefficient, $D_{\rm
lys}=13.7\times 10^{-7}$cm$^2$/sec, which was measured for
lysozyme \cite{mattisson00}.  The Einstein relation for the diffusion
coefficient together with the Stokes law yield $D=\kappa_B T/6\pi\eta
r$ where $K_B$ is the Boltzmann constant, $r$ is the radius of the
diffusing object, and $\eta$ is the viscosity. The latter can be
evaluated through kinetic theory as $\eta\sim n \kappa_B T \tau$,
where $n$ is the density of the viscous medium in which diffusion
takes place and $\tau$ is the mean flight time between collision with
solvent molecules setting the time scale \cite{chaikin95}. The
resulting expression for the diffusion coefficient $D\sim 1/n r\tau$
allows us to get $\frac{\tau_{\rm lys}}{\tau_{\rm pep}}=\frac{D_{\rm
pep}r_{\rm pep}}{D_{\rm lys}r_{\rm lys}}\simeq0.195$ picoseconds as an
estimate of the real time corresponding to one collision step in our
molecular dynamics simulations.  We use $r_{\rm lys}=19\AA$ as an
estimate of $r$ for lysozyme, whereas we take $r_{\rm pep}=5.85\AA$ as
the average radius of gyration of the peptide as found in our
simulations.  Hence, the total number of collision steps, $4\times
10^9$, performed in every simulation corresponds qualitatively to
$0.78$ milliseconds.

%\subsection*{Animations}
%Animation one (Video S1, Supplementary Material)
%shows configurations obtained from the molecular dynamics 
%trajectory that corresponds to Fig.~\ref{fig1}.
%Animation two (Video S2, Supplementary Material) 
%shows the final configuration obtained from the 
%molecular dynamics trajectory shown to Fig.~\ref{fig1}.

%\bibliographystyle{plos2009}
%\bibliography{biblio_title}

\section*{Acknowledgements} 
We thank Sara Linse, Amos Maritan, Flavio Seno and Wei-Feng Xue for
enlightening discussions. 
%This work was supported by the University
%of Padua through Progetto di Ateneo N. CPDA083702 and through PRIN
%Project 2007 N. 2007B57EAB.

\clearpage

\clearpage

\clearpage

\clearpage

\clearpage

\begin{table}
\begin{scriptsize}
\begin{center}
\begin{tabular}{|c|c|c||c|c|c||c|c|c||c|c|c||c|c|c|} \hline
  \multicolumn{3}{|c||}{$\alpha$ helix}  & \multicolumn{3}{c||}{$\beta$ (parallel)} & \multicolumn{3}{c||}{$\beta$ (anti-parallel $1$)} & 
    \multicolumn{3}{c||}{$\beta$ (anti-parallel $2$)} & \multicolumn{3}{c|}{co-operativity} \\ \hline     
  $C_{\alpha}$  & $d_1$ & $d_2$ & $C_{\alpha}$  & $d_1$ & $d_2$ & $C_{\alpha}$  & $d_1$ & $d_2$ & $C_{\alpha}$   & $d_1$ & $d_2$ & $C_{\alpha}$  & $d_1$ & $d_2$\\ \hline
  (i,i+3)  & 4.75 & 5.6  & (i,j)     & 4.3  & 5.5  & (i,j)     & 3.95 & 4.8  & (i,j)     & 4.9  & 5.7 & (i-1,i+1)& 5.6  & 7.4 \\ \cline{1-15}
  (i,i+2)  & 5.1  & 5.7  & (i,j-1)   & 5.3  & 6.85 & (i,j-1)   & 4.95 & 6.25 & (i,j-1)   & 5.05 & 6.3 & (i,i+2)  & 5.65 & 7.5 \\ \cline{1-15}
  (i+1,i+3)& 5.2  & 5.75 & (i-1,j)   & 5.2  & 6.95 & (i-1,j)   & 5.0  & 6.4  & (i-1,j)   & 5.1  & 6.5 & (i-1,i+2)& 9.5  & 10.8\\ \cline{1-15}
  (i-1,i+3)& 5.8  & 6.7  & (i+1,j+1) & 4.1  & 5.6  & (i+1,j-1) & 4.7  & 6.0  & (i+1,j-1) & 3.8  & 5.25& (j-1,j+1)& 5.6  & 7.4 \\ \cline{1-15}
           &      &      & (i-1,j-1) & 4.2  & 5.65 & (i-1,j+1) & 4.75 & 6.1  & (i-1,j+1) & 3.85 & 5.35& (j,j+2)  & 5.65 & 7.5 \\ \cline{1-15}
           &      &      & (i+1,j-1) & 6.15 & 9.9  &           &      &      &           &      &     & (j-1,j+2)& 9.5  & 10.8\\ \cline{1-15}
\end{tabular}
\caption{{\bf Summary of distances used to define the various hydrogen
  bonds.} Here $d_1$ and $d_2$ are the lower and the upper threshold
  distances obtained from Fig.~\ref{Fig4} or from similar
  histograms. The hydrogen bonds are formally assigned to $C_{\alpha}$
  atoms $i,i+3$ ($\alpha$-helix), $i,j$ (parallel and anti-parallel
  $\beta$-sheet), and the cooperative hydrogen bonds are formed
  between the pairs $(i,j)$ and $(i+1,j+1)$ for parallel
  $\beta$-sheets or between the pairs $(i,j)$ and $(i+1,j-1)$ for
  anti-parallel $\beta$-sheets.\label{Tab1}}
\end{center}
\end{scriptsize}
\end{table}

\end{document}